\documentclass[%
reprint,
superscriptaddress,
showpacs,
amsmath,amssymb,
aps,
pra,
longbibliography,
]{revtex4-2}

\usepackage{hyperref}
\usepackage{setspace}
\usepackage{color}
\usepackage{graphicx}
\usepackage{dcolumn}
\usepackage{bm}
\usepackage{amsmath}
\usepackage{indentfirst} 
\usepackage{float}
\usepackage{ulem}
\usepackage{tikz}
\usepackage{lineno}

\begin{document}

\title{ Vortex photoelectron holography in strong-field tunneling ionization }

\author{Yongkun Chen}
\affiliation{School of Physics and Wuhan National Laboratory for Optoelectronics, Huazhong University of Science and Technology, Wuhan 430074, China}

\author{Oleg I. Tolstikhin}
\email{Contact author: tolstikhin.oi@mipt.ru}
\affiliation{Moscow Institute of Physics and Technology, Dolgoprudny 141700, Russia}

\author{Toru Morishita}
\email{Contact author: toru@pc.uec.ac.jp}
\affiliation{Institute for Advanced Science, The University of Electro-Communications, 1-5-1 Chofu-ga-oka, Chofu-shi, Tokyo 182-8585, Japan}

\author{Yueming Zhou}
\email{Contact author: zhouymhust@hust.edu.cn}
\affiliation{School of Physics and Wuhan National Laboratory for Optoelectronics, Huazhong University of Science and Technology, Wuhan 430074, China}
\affiliation{Hubei Fundamental Research Center for Physics, Wuhan, 430074, China}

\author{Peixiang Lu}
\affiliation{School of Physics and Wuhan National Laboratory for Optoelectronics, Huazhong University of Science and Technology, Wuhan 430074, China}

\begin{abstract}

Vortex electrons, characterized by a helical phase front, offer unique advantages 
for probing material structures. Such electrons can be 
generated via tunneling ionization in strong laser fields. 
In this paper, we investigate the rescattering dynamics 
of vortex photoelectrons by the parent ion. Specifically, we introduce vortex photoelectron holography, extending conventional strong-field photoelectron holography (SFPH) from plane-wave to vortex rescattering. 
By solving the time-dependent Schr\"{o}dinger equation, we extract the vortex 
scattering phase from the SFPH fringes, showing excellent agreement with scattering 
calculations. Thus, our work provides direct access to the vortex scattering phase, 
paving the way for applying SFPH to structurally sensitive imaging with 
phase-engineered photoelectrons.

\end{abstract}

\maketitle

Vortex electrons, characterized by their unique phase structure, have attracted significant attention over the past decade \cite{BLIOKH20171,RevModPhys.89.035004,IVANOV2022103987}. The vortex electron possesses a helical phase front of the form $e^{im\varphi}$, where $m$ is the 
projection of its orbital angular momentum along the vortex axis. This phase wraps around a central singularity located on the vortex axis, where the 
electron probability density vanishes. This intrinsic structure distinguishes vortex 
electrons from conventional plane-wave electrons, endowing them with unusual 
behaviors and phenomena when interacting with light or matter. For example, the presence of the phase singularity makes the scattering process of vortex electrons extremely sensitive to the distance $b$ between the vortex axis and the scattering center \cite{PhysRevA.92.012705,PhysRevA.98.022706,PhysRevA.98.042701,%
PhysRevA.107.053114,PhysRevA.111.052810,Strnat_2025,Harris_2025}. Since their experimental realization \cite{Uchida2010,Verbeeck2010}, vortex electrons have been widely applied in various fields, such as electron diffraction \cite{PhysRevLett.134.073001}, electron microscopy \cite{PhysRevLett.113.066102,PhysRevLett.116.127203}, and nanoparticle manipulation \cite{2013How}, as well as used as probes for discriminating the chirality of optical near-fields \cite{PhysRevApplied.22.054017}, nanomaterials \cite{harvey2015}, and molecules \cite{gzzh-7r5m}. The successful manipulation and application of free vortex electrons naturally stimulates interest in extending such investigations to more extreme environments. 

Recent studies have demonstrated that, in the strong-field regime vortex 
photoelectrons can be generated either by absorbing 
photons from an optical vortex \cite{PhysRevA.110.L031101,jtxw-363p} or via 
ionization from a bound state with a helical phase distribution \cite{PhysRevA.99.063415}. Especially, vortex photoelectron generation is very efficient in chiral molecules \cite{PhysRevA.110.033107, photom,9t4d-xj5v}, and chiral discrimination can be achieved by measuring the helical phase of the photoelectrons with various techniques \cite{PhysRevLett.129.233201,gcxt-18gk}. In strong-field ionization, a fraction of these liberated photoelectrons can be driven back by the laser field and undergo rescattering by the parent ion. During this process, the laser field controls the relative position between the vortex axis of the returning photoelectron and the scattering center. For instance, in the case of linearly polarized light, the parent ion is located on the vortex axis, resulting in a vanishing impact parameter $b=0$. This built-in alignment eliminates the spatial averaging effects that typically arise from the random distribution of scattering centers in the interaction between external vortex electron beams and mesoscopic targets 
\cite{PhysRevA.98.042701,PhysRevA.92.012705,PhysRevA.98.022706}. This feature establishes vortex photoelectron rescattering in strong-field ionization as a powerful and sensitive spectroscopic tool for probing atomic and molecular structures \cite{PhysRevA.99.063415}.
  

As a well-established technique based on electron rescattering, strong-field 
photoelectron holography (SFPH) \cite{doi:10.1126/science.1198450,Figueira_2020} 
has been extensively utilized to probe atomic and molecular structures, as well as 
ultrafast electron dynamics \cite{Meckel2014,PhysRevLett.116.163004,PhysRevLett.121.253203,PhysRevLett.127.263202,Porat2018,PhysRevLett.122.183202,PhysRevLett.120.133204,doi:10.34133/2022/9842716}. 
This technique relies on the interference between directly tunneled electrons and 
near-forward rescattered electrons, thereby encoding the scattering phase into the 
holographic fringes of the 
photoelectron momentum distribution (PEMD) \cite{PhysRevLett.116.173001}. The retrieval of 
the photoelectron scattering phase from SFPH has been established theoretically 
\cite{PhysRevLett.116.173001} and recently demonstrated experimentally \cite{wcl3-x52t}. 
However, these studies considered only the rescattering of plane-wave photoelectrons. 
Furthermore, although the scattering of external free vortex electrons 
\cite{PhysRevA.89.032715,PhysRevA.91.032703,PhysRevD.94.076001,IVANOV2022103987,Strnat_2025,PhysRevA.98.042701,PhysRevA.98.022706,PhysRevA.92.012705,PhysRevA.111.052810,Harris_2025} 
has been extensively investigated, and the rescattering of vortex photoelectrons 
in a strong laser field \cite{PhysRevA.99.063415,PhysRevA.107.053114} has 
attracted increasing attention, their scattering phase remains unexplored. In this 
paper, we introduce vortex photoelectron holography and demonstrate the extraction 
of the vortex photoelectron scattering phase from the interference fringes. Our 
results provide direct access to this scattering phase and contribute to a deeper 
understanding of the ultrafast dynamics of vortex photoelectrons in strong-field 
ionization.

\begin{figure*}
	\centering
	\includegraphics[width=0.9\linewidth]{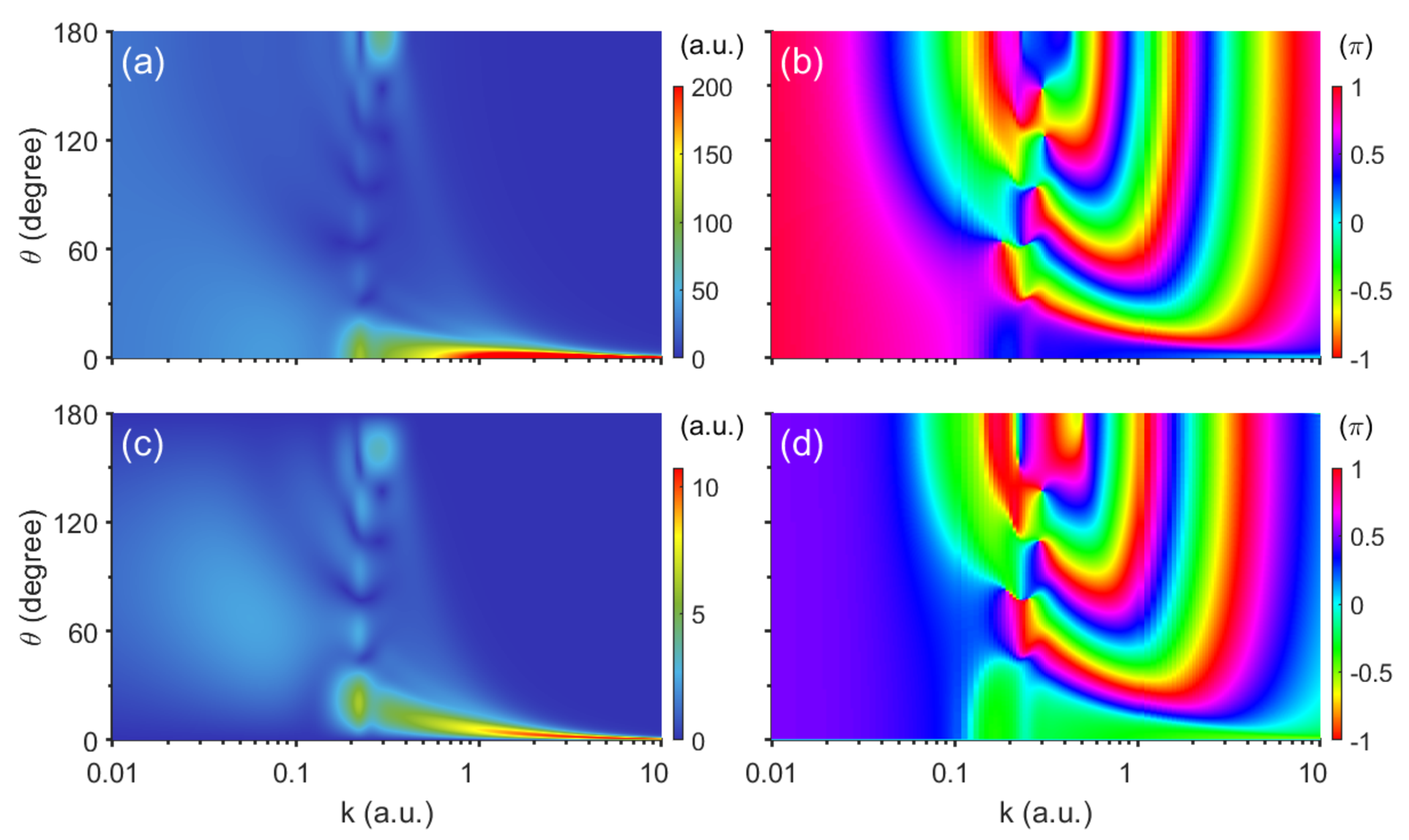}
	\caption{ (a) Absolute value and (b) phase of the plane-wave scattering amplitude $f_k(\theta)$ for the potential (\ref{pot}) with $a=15$
    as functions of the incident momentum $k$ and the scattering angle $\theta$. (c), (d) Same as in (a), (b), respectively, but for the vortex scattering amplitude $f_{mk\theta_k}(\theta)$ with $m=1$.}
	\label{sca_amp}
\end{figure*} 

We illustrate our results by calculations for a finite-range potential (atomic units are used unless otherwise stated),
\begin{equation}	
    V(r)=-Z\exp[-(r/a)^2]/r,
    \label{pot}
\end{equation}
where $Z=2$ is chosen to model He$^+$. The Gaussian screening
factor is introduced to eliminate the effect of the long‑range Coulomb tail.
Let us first compare the scattering amplitudes for plane-wave and vortex electrons. The plane-wave scattering states satisfy the asymptotic boundary condition \cite{1982Physics}
\begin{equation}	
    \psi_\mathbf{k}(\mathbf{r}) \Big|_{r \rightarrow \infty } = e^{i\mathbf{k} \cdot \mathbf{r}} + f_k(\theta)\frac{e^{ikr}}{r} ,
    \label{psip}
\end{equation}
where $\mathbf{k}$ is the incident electron 
momentum and $f_k(\theta)$ is the plane-wave scattering amplitude. 
Vortex electrons produced by tunneling ionization
in a linearly polarized laser field return for rescattering
with a zero impact parameter, that is, their vortex axis coincides 
with the $z$ axis \cite{PhysRevA.99.063415}. 
In this case, the vortex scattering states have 
the asymptotic form
\begin{align}
    \psi_{mk\theta_k}(\mathbf{r}) \Big|_{r \rightarrow \infty } = & \left[
    J_{|m|}(k_\perp r_\perp)e^{ ik_zz} 
    \right. \nonumber \\ & \left.
    + f_{mk\theta_k}(\theta)\frac{e^{ikr}}{r}\right]e^{im\varphi}.
    \label{psiv}
\end{align}
Here, 
$k_\perp=k\sin\theta_k$ and $k_z=k\cos\theta_k$ are the transverse and 
longitudinal components of the incident electron momentum,
$r_\perp=\sqrt{x^2+y^2}$, and 
$f_{mk\theta_k}(\theta)$ is the vortex scattering amplitude. 
The vortex scattering state is given by a superposition of plane-wave scattering states whose momenta $\mathbf{k}$ make an 
angle $\theta_k$ with the $z$ axis.
As a result, the vortex scattering amplitude can be expressed in the form 
\begin{equation} 
    f_{mk\theta_k}(\theta) = (-i)^{|m|} \int_0^{2\pi} e^{im\varphi_k} f_k(\theta') \frac{d \varphi_k}{2\pi},
    \label{fv1}
\end{equation}
where the angle $\theta'$ is defined by $\cos\theta' =\sin\theta\sin\theta_k\cos\varphi_k +  
\cos\theta\cos\theta_k$.

In the adiabatic regime, the returning photoelectron has zero transverse 
momentum \cite{PhysRevA.86.043417,PhysRevA.99.063415}, meaning $k_\perp=0$. As $k_\perp$ approaches 
zero, the absolute value of $f_{mk\theta_k}(\theta)$ 
defined in Eq.~(\ref{psiv}) scales as $k_\perp^{|m|}$.
Therefore, in the paraxial limit $k_\perp \to 0$, it is more
convenient to use an alternative definition of the vortex 
scattering amplitude \cite{PhysRevA.99.063415,PhysRevA.107.053114}, in which
the factor $k_\perp^{|m|}$ is canceled out.
However, the phase of $f_{mk\theta_k}(\theta)$ 
becomes independent of $k_\perp$ for sufficiently small values of $k_\perp$.
Since we are primarily 
interested in the phase, we retain the definition in Eq.~(\ref{psiv}) and
set $k_\perp=0.01$ in the following calculations.

\begin{figure}[h]
	\centering
	\includegraphics[width=\linewidth]{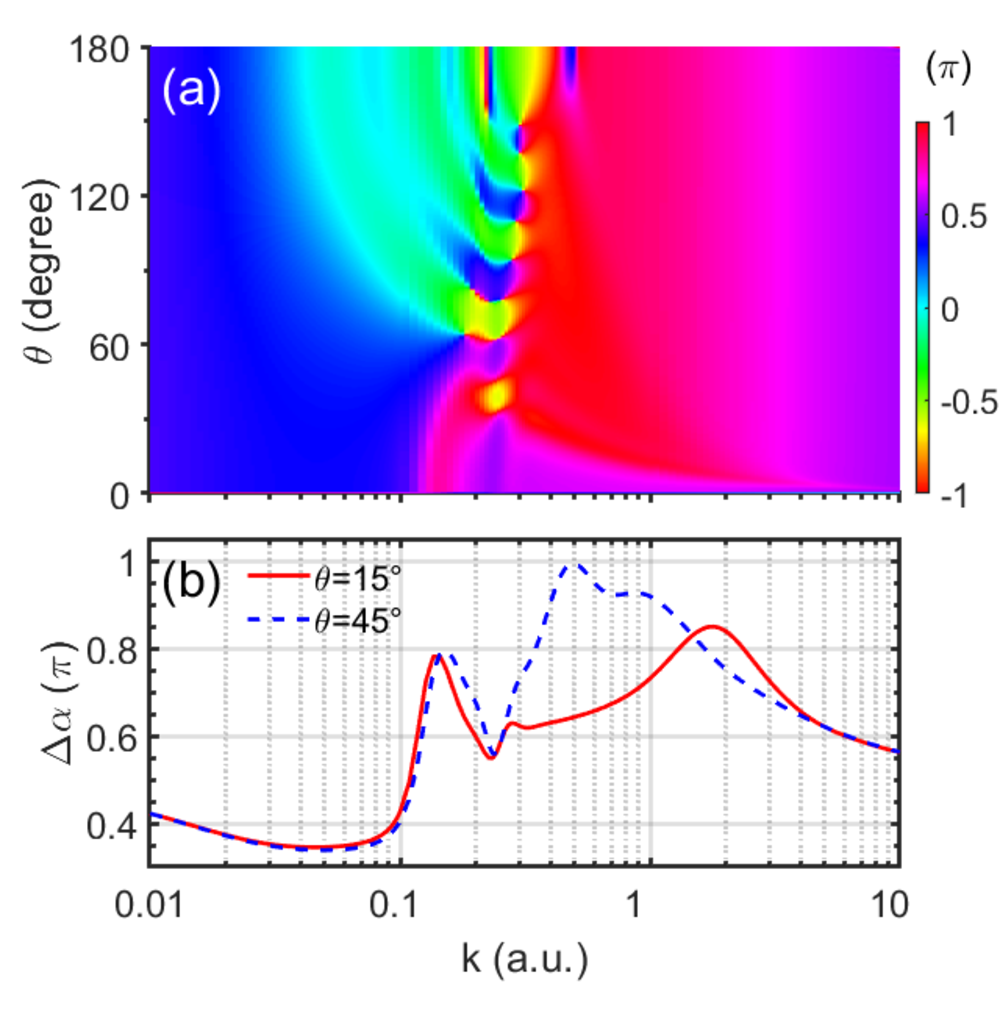}
	\caption{ (a) Difference $\Delta\alpha$ between the phases of the
    plane-wave $f_k(\theta)$
    and vortex $f_{m=1,k\theta_k}(\theta)$ scattering amplitudes shown in Fig.~\ref{sca_amp}.
    (b) Cuts of the scattering phase difference at scattering angles $\theta=15^\circ$ (red solid line) and $\theta=45^\circ$ (blue dashed line).}
	\label{pha_diff}
\end{figure}

Using standard methods 
of scattering theory (ST) \cite{1982Physics}, we calculated the plane-wave 
scattering amplitude for the potential (\ref{pot}) with the screening parameter $a=15$. 
Figures \ref{sca_amp}(a) and \ref{sca_amp}(b) present its absolute value and phase as functions of the incident momentum $k$ and the scattering angle $\theta$. 
A broad $g$-wave resonance is clearly visible at $k\approx0.23$, where the scattering phase varies dramatically. For $k>0.1$, the absolute value of $f_k(\theta)$ peaks sharply in the forward scattering direction at $\theta=0$. The vortex scattering
amplitude is calculated using Eq.~(\ref{fv1}) for the same potential 
and $m=1$. Its absolute value and phase 
are shown in Figs.~\ref{sca_amp}(c) and (d), respectively. 
This amplitude also exhibits the resonance feature. 
Note that, unlike the plane-wave case,  
$f_{mk\theta_k}(\theta)$ exactly vanishes at $\theta=0$. 
For $k>0.1$, its absolute value
peaks near the forward scattering direction. 
The peak is located at $\theta \approx 15^\circ$ for $k=0.3$, 
and this angle gradually decreases, 
approaching zero as the incident momentum increases.
Although the vortex scattering phase exhibits a trend similar 
to that of the plane-wave phase --- namely, 
at a given scattering angle, both phases 
decrease monotonically with increasing incident momentum --- 
their values differ 
considerably.

Figure~\ref{pha_diff} illustrates the phase difference between the plane-wave 
and vortex scattering amplitudes more clearly. At small and large momenta 
outside the range $0.1 \lesssim k \lesssim 1$ dominated by the resonance 
feature, the phase difference depends only weakly on the scattering angle $\theta$, 
approaching a constant as $k \to 0$ and monotonically decreasing for $k \gtrsim 1$. 
However, near the resonance, the phase difference depends strongly on both $k$ 
and $\theta$. Crucially, at momenta $k \sim 1$ of main interest in strong-field 
physics and within the near-forward angular range $\theta \lesssim 30^\circ$, 
the phase difference is large, on the order of $\pi$. Thus, extracting the 
scattering phase from SFPH fringes allows for unambiguous discrimination 
between the rescattering of plane-wave and vortex photoelectrons.

To demonstrate the extraction procedure, we 
numerically solve the three-dimensional time-dependent Schr\"{o}dinger equation 
(TDSE) with the potential (\ref{pot}),
\begin{equation}
\begin{aligned}	
	\label{TDSE}	
    i\frac{\partial\Psi(\mathbf{r},t)}{\partial t}  = \left[ \frac{\hat{\mathbf{p}}^2}{2} + \hat{\mathbf{p}} \cdot \mathbf{A}(t) + V(r) \right] \Psi(\mathbf{r},t).
\end{aligned}
\end{equation}
Here, $\mathbf{A}(t)=-\int\mathbf{E}(t) dt$ is the vector potential of a linearly 
polarized laser field $\mathbf{E}(t)=E(t)\mathbf{e}_z$. As the initial state, 
we use the $2p1$ state of the field-free system. 
The wavefunction is expanded in a partial-wave series 
\begin{equation}
\begin{aligned}
    \Psi(\mathbf{r},t)=\frac{1}{r}
    \sum_{l=0}^{l_\text{max}}R_{l}(r,t)Y_{l1}(\theta,\varphi),
\end{aligned}
\end{equation}
where $R_{l}(r,t)$ are the radial functions and $Y_{l1}(\theta,\varphi)$ are spherical harmonics
with $m=1$. In our calculations, $R_{l}(r,t)$ are represented using the finite-element discrete variable representation method \cite{PhysRevA.62.032706}, with a box size of $r_\text{max}=600$. The results converge at $l_\text{max}=400$. 
The time propagation of the wavefunction is implemented using the split-Lanczos method \cite{Jiang:17} with a time step of $0.01$. In each propagation step, a mask function $F(r)=1-1/[1+e^{(400-r)/2}]$ is used to split the wavefunction into an inner part $\Psi_\text{in}(\mathbf{r},t)=F(r)\Psi(\mathbf{r},t)$ and an outer part $\Psi_\text{out}(\mathbf{r},t)=[1-F(r)]\Psi(\mathbf{r},t)$. The inner part $\Psi_\text{in}(\mathbf{r},t)$ evolves according to the full Hamiltonian, while the outer wavefunction is propagated by a Volkov propagator \cite{PhysRevA.77.013401}. The PEMD is obtained by projecting the outer wavefunction onto the field-free plane-wave scattering states. For the present initial state, the photoelectron released by tunneling is described by a vortex wave packet that carries a helical phase corresponding to $m=1$. This model provides a convenient framework for investigating SFPH with vortex photoelectrons.

\begin{figure}[tb]
	\centering
	\includegraphics[width=\linewidth]{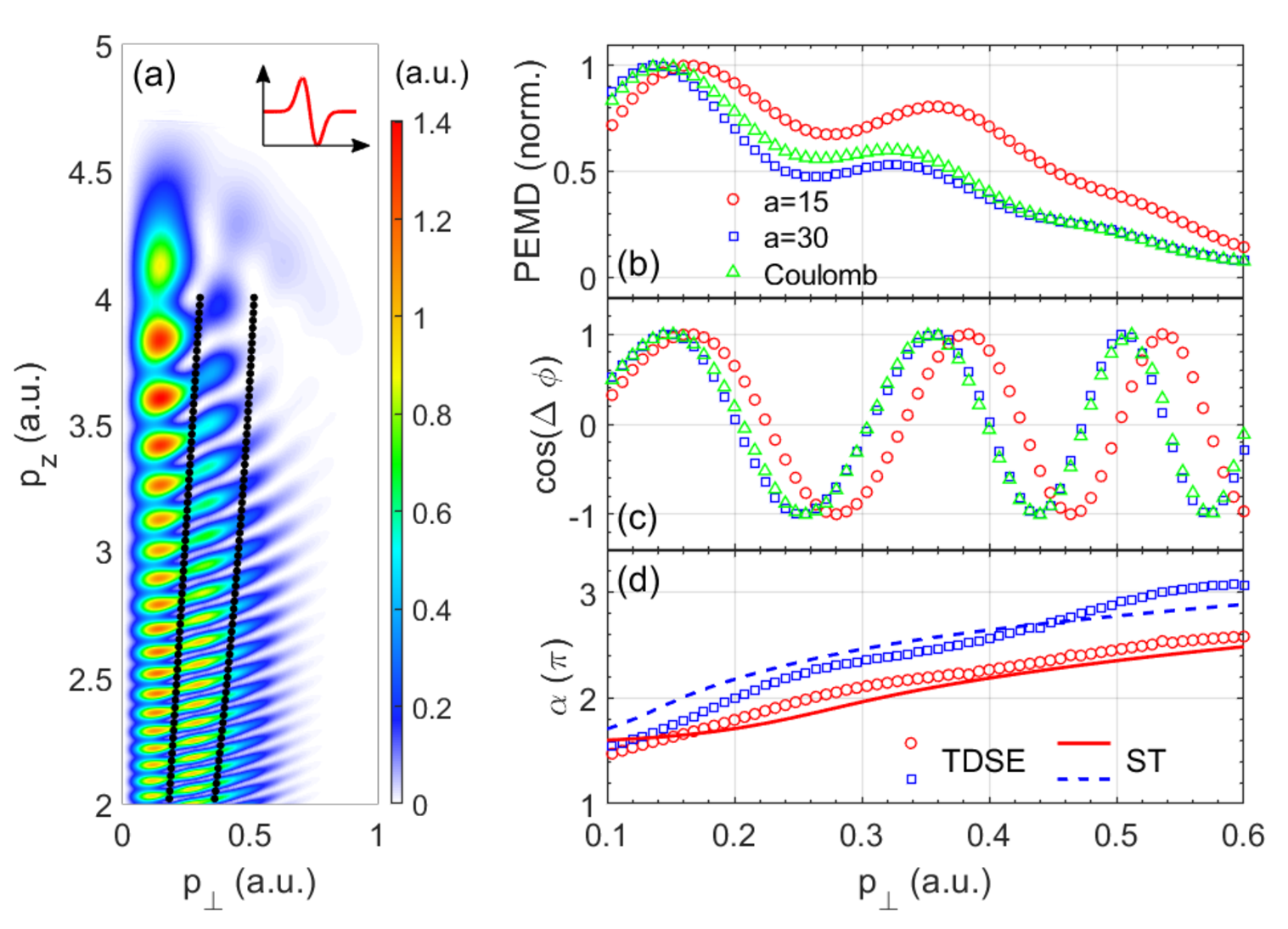}
	\caption{(a) PEMD for ionization from the $2p1$ state in the potential (\ref{pot}) with $a=30$ by a single-cycle pulse. Black dots indicate the minima of the SFPH interference pattern. The inset schematically shows the temporal profile of the laser field. (b) Cuts of the normalized averaged PEMD at $p_z=3$ for potentials with screening parameters $a=15$ (red circles) and $a=30$ (blue squares), and for the pure Coulomb potential (green triangles). (c) Oscillating factor $\cos(\Delta\phi)$ extracted from (b) as a function of $p_\perp$. (d) Scattering phase $\alpha$ retrieved from (c) as a function of $p_\perp$. The TDSE results (symbols) are compared with the ST results (lines).}
	\label{sin_cyc}
\end{figure}

We begin with a single-cycle laser pulse of the form 
$E(t)=-\sqrt{2e}E_0(2t/\tau)\exp[-(2t/\tau)^2]$ 
to reduce the complexity of the interference pattern and highlight the SFPH fringes. 
The PEMD calculated for the potential (\ref{pot}) with the screening parameter 
$a=30$, pulse amplitude $E_0=0.1$, and duration $\tau=75$ (corresponding to an 
intensity of $3.5 \times 10^{14}$~W/cm$^2$ and a wavelength of $\lambda \approx 800$~nm, 
respectively) is shown in 
Fig.~\ref{sin_cyc}(a). The inset shows the temporal profile of $E(t)$. 
The PEMD exhibits two types of interference fringes. The near-horizontal interference 
structure results from the interference between direct electrons liberated during 
two adjacent quarter cycles of the laser pulse \cite{PhysRevA.74.063407,PhysRevA.81.021403,PhysRevLett.114.143001}. 
This rapidly oscillating structure is generally difficult to observe experimentally. 
The other, near-vertical interference structure is SFPH, which is the focus of 
this study. The black dots in 
Fig.~\ref{sin_cyc}(a) indicate the positions of the 
minima of the SFPH fringes. Unlike plane-wave photoelectrons, for which the SFPH 
exhibits a maximum at $p_\perp=0$, the PEMD for vortex photoelectrons vanishes 
along the $p_z$ axis. 
The SFPH fringes originate from the interference between photoelectrons reaching 
the detector directly after tunneling and those undergoing rescattering by the 
parent ion before reaching the detector. Thus, the resulting interference pattern 
encodes the corresponding phase difference. Notably, only photoelectrons rescattered 
in the near-forward direction contribute to the formation of the SFPH fringes. 
According to adiabatic theory, this phase difference can be expressed as \cite{PhysRevA.86.043417,PhysRevLett.116.173001}
\begin{equation}
\begin{aligned}
    \Delta \phi (p_\perp)=\frac{1}{2}(t_r-t_i)p_\perp^2 + \alpha,
    \label{psi}
\end{aligned}
\end{equation}
where $t_i$ and $t_r$ are the ionization and rescattering times, and $\alpha$ is 
the phase of the vortex scattering amplitude $f_{m=1,k\theta_k}(\theta)$. The first 
term in Eq.~(\ref{psi}) corresponds to the phase difference between the direct and 
rescattered photoelectrons, accumulated during propagation in the laser field. 
Given the laser field shape, this term can be readily calculated using classical 
equations of motion \cite{doi:10.1126/science.1198450,PhysRevLett.116.173001}. The second term, $\alpha$, 
is the scattering phase of the photoelectron acquired during rescattering by the 
parent ion. It is this phase that we aim 
to extract from the SFPH fringes.

To do this, we employ the procedure introduced 
in Ref.~\cite{PhysRevLett.116.173001}. First, we average the PEMD over the 
interval $2.9 \leq p_z \leq 3.1$ to eliminate the near-horizontal interference. 
The resulting averaged PEMD at $p_z=3$, normalized to unity at its maximum, 
is shown in Fig.~\ref{sin_cyc}(b). The results for the potential (\ref{pot}) 
with screening parameters $a=15$ and $a=30$ are shown by red circles and blue 
squares, respectively. 
We also present the results for the pure Coulomb potential $V(r)=-Z/r$ 
($a=\infty$) using green triangles.
Next, we remove a smooth envelope of the PEMD and obtain the oscillating factor 
$\cos(\Delta\phi)$ bounded between $-1$ and $1$, where $\Delta\phi$ is the 
interference phase. This factor is shown in Fig.~\ref{sin_cyc}(c)
for the two values of $a$ as well as for the pure Coulomb potential. Finally, 
by equating the resulting $\Delta\phi$ to the phase given by Eq.~(\ref{psi}), 
we extract the vortex scattering phase $\alpha$. This phase as a function of 
$p_\perp$, calculated for screening parameters $a=15$ and $a=30$, is shown in 
Fig.~\ref{sin_cyc}(d). For both values of $a$, the TDSE and ST results agree 
well with each other.

\begin{figure}[tb]
	\centering
	\includegraphics[width=\linewidth]{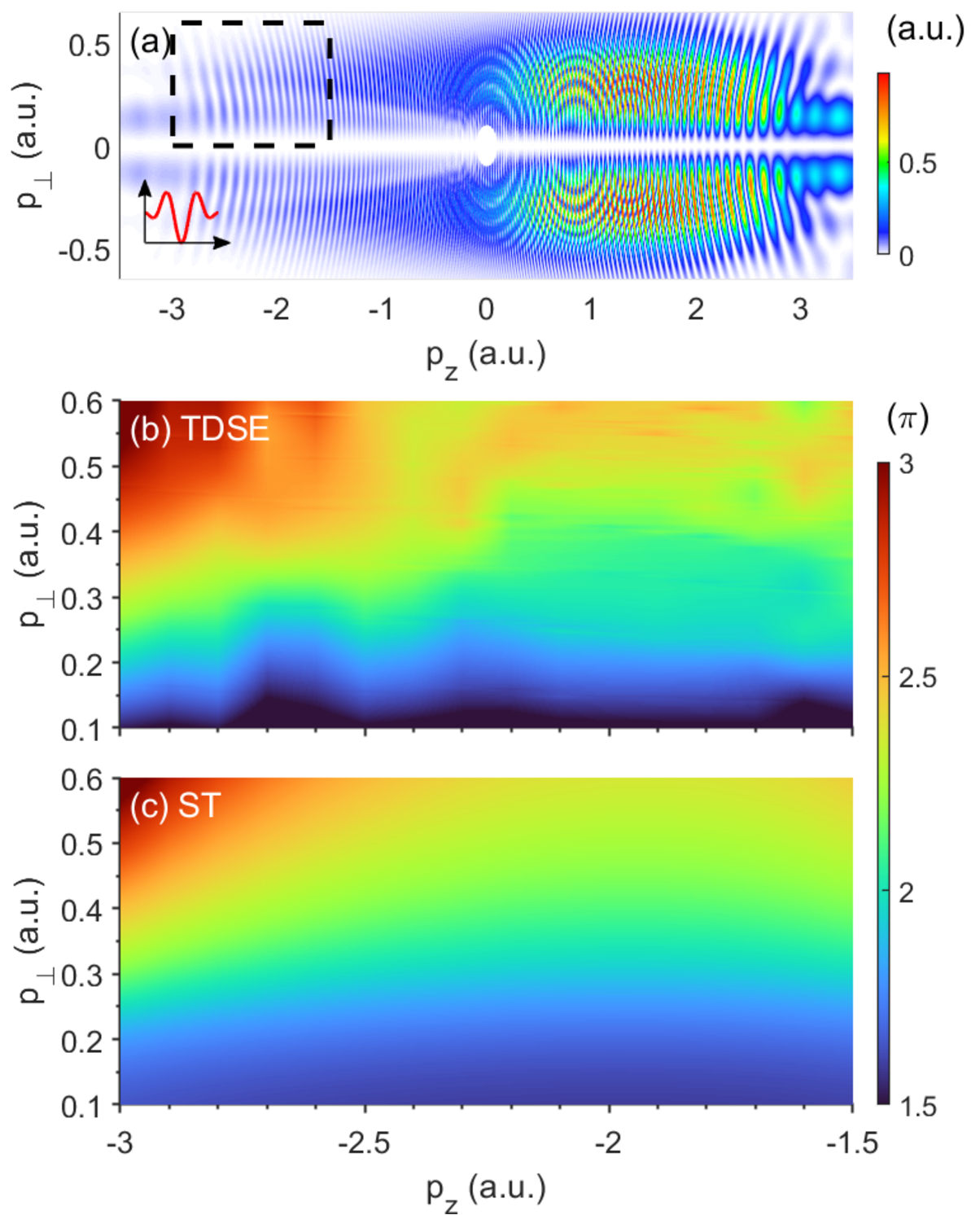}
	\caption{ (a) PEMD for ionization from the $2p1$ state in the potential (\ref{pot}) with $a=15$ by a few-cycle pulse. Inset: schematic of the laser field. (b) Phase $\alpha$ of the vortex scattering amplitude $f_{m=1,k\theta_k}(\theta)$ extracted by solving the TDSE from the region of the PEMD indicated by the dashed rectangle in (a). (c) Same as (b), but calculated using ST. }
	\label{few_cyc}
\end{figure}

As the screening parameter increases, the interference phase $\Delta\phi$ 
converges to that for the Coulomb potential. Indeed, as shown in Fig.~\ref{sin_cyc}(c), the oscillating factor $\cos(\Delta\phi)$ obtained for $a=30$ almost coincides with that for the Coulomb case. However, Eq.~(\ref{psi}) can not apply to the pure Coulomb potential. This limitation arises because this equation is derived from the adiabatic theory for finite-range potentials, which does not account for 
the long-range Coulomb tail \cite{PhysRevA.86.043417,PhysRevLett.116.173001}. 
Despite this limitation, the scattering phase retrieved using Eq.~(\ref{psi}) is determined by the behavior of the potential within a finite region. As demonstrated in Ref.~\cite{wcl3-x52t}, it encodes valuable information about the 
target structure.

We next consider a more realistic few-cycle laser pulse with 
$\mathbf{A}(t)=A_0\cos^2(\pi t/2T)\sin(\omega t)\mathbf{e}_z$, where $A_0=4.18$ 
(an intensity of $5 \times 10^{14}$ W/cm$^2$), $\omega=0.0285$ (a wavelength of 
$1600$~nm), and $T=2\pi/\omega$. 
The PEMD calculated with $a=15$ is shown in Fig.~\ref{few_cyc}(a). The inset 
shows the temporal profile of the field. We select the $p_z$ range from $-3$ 
to $-1.5$ [as indicated by the dashed rectangle in Fig.~\ref{few_cyc}(a)], 
where the holographic fringes are most pronounced. For each $p_z$ in this range, 
we apply the previously described procedure to the cut of the PEMD along 
$p_\perp$, yielding the scattering phase $\alpha$ as a function of $p_\perp$ 
and $p_z$, as shown in Fig.~\ref{few_cyc}(b). Figure~\ref{few_cyc}(c) presents 
the corresponding results calculated using ST. As can be seen, the TDSE and ST 
results agree well across this wide $(p_z,p_\perp)$ region.

To realize the proposed scheme experimentally, one can employ circularly polarized 
light to excite the He$^+$ electron from the $1s$ to the $2p1$ state, without 
populating the degenerate $2p0$ state. Although some population will remain in 
the $1s$ state, its large binding energy suppresses tunneling ionization from 
this state. Consequently, when an intense laser pulse subsequently drives tunneling 
ionization, only vortex photoelectrons carrying a helical phase are liberated from 
the $2p1$ state and contribute to the SFPH signal, without any plane-wave photoelectron 
admixture.

In summary, we have studied vortex photoelectron holography in strong-field tunneling ionization. In laser-field induced rescattering, the impact parameter is controllable, allowing the scattering center to be easily positioned at the phase singularity of vortex photoelectron without the ion trap \cite{Schmiegelow2016,PhysRevLett.129.263603}. This ability to align the phase singularity of a vortex electron with the scattering center circumvents the spatial averaging inherent in the conventional external vortex electron scattering scheme, enabling site-specific access to the target structure. The extracted vortex scattering phase is found to be in excellent agreement with independent scattering calculations. Although the present demonstration was performed with atomic photoionization, the method can be naturally extended to molecular systems. In polyatomic molecules, the scattering of vortex photoelectrons is highly sensitive to the position of each atom due to the phase singularity \cite{PhysRevA.98.042701,PhysRevA.98.022706,PhysRevA.92.012705,Strnat_2025,PhysRevA.111.052810,Harris_2025,PhysRevA.107.053114}, making them an ideal tool for probing the molecular structure. More broadly, this work generalizes laser-induced photoelectron holography from plane-wave to vortex beams, introducing an additional degree of freedom, orbital angular momentum, into ultrafast scattering imaging. Beyond the present demonstration, this capability holds particular promise for detecting molecular chirality and resolving transient structural rearrangements on attosecond timescales.

\section*{acknowledgments}
We thank S. Furuhata for his help with calculations in the early stage of the work. This work was supported National Natural Science Foundation of China (Grants No. U25D8005, 12374264, 12547161), National Key Research and Development Program of China (Grant No. 2023YFA1406800), Basic Research Support Program of Huazhong University of Science and Technology (2024BRA002). The computing work in this paper is supported by the Public Service Platform of High Performance Computing provided by Network and Computing Center of HUST.
O.~I.~T. acknowledges support from the Ministry of Science
and Higher Education of the Russian Federation 
(Grant No.~FSMG-2026-0012).

\bibliography{references}

\end{document}